\documentclass[aps,prl,twocolumn,amsmath,amssymb,superscriptaddress,showpacs]{revtex4}
\usepackage{graphicx}
\usepackage{hyperref}
\usepackage{bm}
\usepackage{multirow}
\usepackage{times,color}

\begin{document}

\title{Von Neumann Entropy Spectra and Entangled Excitations in Spin-Orbital Models }

\author {     Wen-Long You }
\affiliation{ Max-Planck-Institut f\"ur Festk\"orperforschung,
              Heisenbergstrasse 1, D-70569 Stuttgart, Germany }
\affiliation{School of Physical Science and Technology, Soochow University,
             Suzhou, Jiangsu 215006, People's Republic of China  }

\author {     Andrzej M. Ole\'s }
\affiliation{ Max-Planck-Institut f\"ur Festk\"orperforschung,
              Heisenbergstrasse 1, D-70569 Stuttgart, Germany }
\affiliation{ Marian Smoluchowski Institute of Physics, Jagellonian
              University, Reymonta 4, PL-30059 Krak\'ow, Poland }

\author {     Peter Horsch }
\affiliation{ Max-Planck-Institut f\"ur Festk\"orperforschung,
              Heisenbergstrasse 1, D-70569 Stuttgart, Germany }

\date{\today}

\begin{abstract}
We consider the low-energy excitations of one-dimensional
spin-orbital models which consist of spin waves, orbital waves, and
joint spin-orbital excitations. Among the latter we identify
strongly entangled spin-orbital bound states which appear as peaks in
the von Neumann entropy (vNE) spectral function introduced in this
work. The strong entanglement of bound states is manifested by a
universal logarithmic scaling of the vNE with system size,
while the vNE of other spin-orbital excitations saturates.
We suggest that spin-orbital entanglement can be experimentally
explored by the measurement of the dynamical spin-orbital correlations
using resonant inelastic x-ray scattering, where strong spin-orbit
coupling associated with the core hole plays a role.
\end{abstract}


\maketitle

{\it Introduction.---}The spin-orbital
interplay is one of the important topics in the theory of strongly
correlated electrons \cite{Tok00}. In many cases, the intertwined
spin-orbital interaction is decoupled by mean-field approximation,
and the spin and orbital dynamics are independent from each other.
Thus a spin-only Heisenberg model can be derived by averaging
over the orbital state, which successfully explains
magnetism and optical excitations in some materials, for instance in
LaMnO$_3$ \cite{Fei99}. But in others, especially in $t_{2g}$ systems
\cite{Kri12}, the orbital degeneracy plays an indispensable role in
understanding the low-energy properties in the Mott insulators of
transition metal oxides (TMOs), such as LaTiO$_3$ \cite{Kha00},
LaVO$_3$ and YVO$_3$ \cite{Hor08},
and also in recently discussed RbO$_2$ \cite{Woh11}. The well
known cases are also strong spin-orbit coupling which leads to locally
entangled states \cite{Jac09}, and entanglement on the superexchange
bonds in K$_3$Cu$_2$F$_7$ \cite{Brz11}. For such
models, the mean-field-type approximation
and the decoupling of composite spin-orbital correlations
fail and generate uncontrolled errors,
even when the orbitals are polarized \cite{Wohlfeld}.
The strong spin-orbital fluctuations
on the exchange bonds will induce the violation of the
Goodenough-Kanamori rules \cite{Andrzej}. Furthermore, the
flavors may form exotic composite spin-orbital excitations.

{\it Model and system.---} A paradigmatic model derived for a TMO
in Mott-insulating limit is the one-dimensional (1D) spin-orbital
Hamiltonian, which reads
\begin{eqnarray}
H &=& -J \sum_\textbf{i} 
\left( \vec{S}_\textbf{i} \cdot\vec{S}_{\textbf{i}+1} + x\right)
\left(\vec{T}_\textbf{i} \cdot \vec{T}_{{\bf i}+1}+y\right),
\label{SO-Hamiltonian}
\end{eqnarray}
where $\vec{S}_\textbf{i}$ and $\vec{T}_\textbf{i}$ are spin-1/2 and
pseudospin-1/2 operators representing the spin and orbital degrees
of freedom located at site $\textbf{i}$, respectively,
and we set below $J=1$.
It is proposed that ultracold fermions in zig-zag optical lattices can
reproduce an effective spin-orbital model \cite{Gsun}. For general
$\{x,y\}$, the model (\ref{SO-Hamiltonian}) has an
SU(2)$\otimes$SU(2) symmetry. An additional $\mathbf{Z}_2$
bisymmetry occurs by interchanging spin and orbital operators when
$x=y$. In the case of $x=y=\frac14$, Hamiltonian (\ref{SO-Hamiltonian})
reduces to a SU(4) symmetric model, which is exactly soluble by the
Bethe ansatz \cite{YouQuanLi,Sutherland}.
There are three Goldstone modes corresponding to separate spin and
orbital excitation, as well as composite spin-orbital excitations in
case of $J<0$, in contrast to a quadratic dependence of the energy
upon the momentum in the long-wave limit for $J>0$. The spectra of
elementary excitations are commonly not analytically soluble away
from the SU(4) point. We will, however, show that the low-energy
excitations can be analytically obtained in some specific phases in
the case when $J>0$, and this offers a platform to study the
spin-orbital entanglement. In this Letter, we
go beyond the ideas developed for spin systems \cite{Alb09}.
We demonstrate that spin-orbital entanglement entropy
clearly distinguishes weakly correlated spin-orbital excitations from
bound states and resonances by its magnitude and distinct scaling behavior.
We propose how to connect the entanglement entropy with experimentally
observable quantities of recently developed spectroscopies.

{\it von Neumann entropy.---} Currently, concepts from quantum
information theory are being studied with the aim to explore
many-body theory from another perspective and vice versa. A
particularly fruitful direction is using quantum entanglement to
shed light on exotic quantum phases \cite{Hui08,Yao10}.
Entanglement entropy even distinguishes phases
in the absence of conventional order parameters \cite{Kit06}.
In general a many-body quantum system is subdivided into $A$ and $B$
parts and the entanglement entropy is the von Neumann entropy (vNE),
${\cal S}_{\rm vN}= -\textrm{Tr}\{\rho_A \log_2 \rho_A \}$, where
$\rho_A=\textrm{Tr}_B\{\rho\}$ is the reduced density matrix of the
subspace $A$ and $\rho$ is the full density matrix. The vNE is bounded,
${\cal S}_{\rm vN} \le \log_2 \textrm{dim}\rho_A$, and
easy to calculate. Experimental determination appears harder, yet
there are proposals involving transport measurements in quantum
point contacts \cite{Klich}.

Interestingly the vNE scales proportionally to the boundary of the
subregion obtained by the spatial partitioning \cite{Eisert}. The
dependence of the boundary or area law can be traced back to study
of black hole physics \cite{Bombelli} and was extensively exploited
for 1D spin chains \cite{Amico}. If the block $A$ is of length $l$
in a system of length $L$ with periodic boundary condition, the vNE
of gapped ground states is bounded as ${\cal S}_l={\cal O}(1)$,
while a logarithmic scaling
${\cal S}_l= c\log_2 l  + {\cal O}(1)$ $(L\gg l\gg 1)$
has been proven to be universal property of the gapless phases in
critical systems by the underlying conformal field theory
\cite{Holzhey}. A violation of the area law is expected for
the low-lying excited states of critical chains \cite{Masanes}. To
date, measurements of the vNE for subdivision of degrees of freedom
other than in spatial segmentation have not been fully explored. 
In a composite system containing spin and orbital operators,
the decomposition of different flavors retains the real-space
symmetries.

\begin{figure}[t!]
\begin{center}
\includegraphics[width=8.0cm]{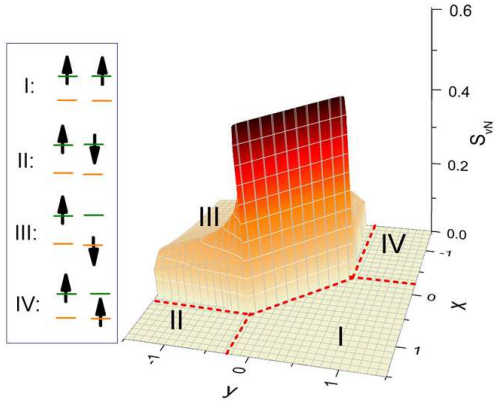}
\caption{(color online). Spin-orbital entanglement ${\cal S}_\textrm{vN}$
in the ground-state of the spin-orbital model (\ref{SO-Hamiltonian}) as a
function of $x$ and $y$ and system size $L=8$. The (red) dashed lines 
mark the critical lines determined by the fidelity susceptibility 
(see text). The two-site configurations in phases I-IV are shown on 
the left. The two orbitals per site are degenerate (their splitting is
only for clarity of presentation).}
\label{Phasediagramoftwosites}
\end{center}
\end{figure}

{\it Phase diagram.---}
A quantum phase transition (QPT) is identified as a
point of nonanalyticity
of the ground state and associated expectation values
in the thermodynamic limit. To shed light on the phase boundaries,
we first consider two sites \cite{Brink},
$H_{12}=- \frac14 (\vec{S}_{12}^2 -\vec{S}_{1}^2-\vec{S}_{2}^2 + 2 x)
(T_{12}^2 -\vec{T}_{1}^2 -\vec{T}_{2}^2  + 2y)$,
where
$\vec{S}_{12}=\vec{S}_{1}+ \vec{S}_{2}$ and
$\vec{T}_{12}=\vec{T}_{1}+ \vec{T}_{2}$.
A pair of spins (orbitals) can form either a singlet with $S_{12}=0$
($T_{12}=0$) or a triplet with $S_{12}=1$ ($T_{12}=1$), and various
combinations of quantum numbers correspond to different phases
shown in Fig. \ref{Phasediagramoftwosites}. In phase I, the state
with $S_{12}=1=T_{12}$ has the lowest energy,
and thus the energy per bond is $e_B^I \ge e_{xy}= - (x+1/4)(y+1/4)$.
For a larger system with $L$ bonds, we have $E_0^I(H) \ge L e_{xy}$.
On the other hand, taking a ferro-ferro state
$\vert 0 \rangle$ as a variational state, $E_0^I(H) \le  L e_{xy}$.
Therefore, the energy of phase I is exactly $E_0^I(H)= L e_{xy}$
and the ferro-ferro state is the corresponding ground state.

Without prior knowledge of order parameter, various characterizations
from the perspective of quantum information theory can be used to
identify phase boundaries. One often used tool is the vNE \cite{YanChen}.
Tracing orbital degrees of freedom, we obtained the spin-orbital vNE
${\cal S}_\textrm{vN}$ for the ground state of $L=8$ chain
in the Hilbert subspace of $S_z=T_z=0$ \cite{YanChen}.
However, here we find that the vNE of the ground state does not
distinguish phase I from phase II or IV --- all three phases
having ${\cal S}_\textrm{vN}=0$
(see Fig. \ref{Phasediagramoftwosites}).
Therefore we use the quantum fidelity to quantify the phase diagram
\cite{IntJModPhys.24.4371}. The fidelity defined as follows,
${\cal F} (\lambda,\delta\lambda)=
|\langle\Psi_0(\lambda)|\Psi_0(\lambda+\delta\lambda)\rangle|$,
is taken along a certain path $\{x(\lambda),y(\lambda)\}$ and reveals
all phase boundaries. The fidelity susceptibility,
$\chi_{\textrm F}\equiv-(2\ln{\cal F}) /(\delta\lambda)^2
\vert_{\delta\lambda\rightarrow 0}$,
exhibits a peak at the critical point, and
can be treated as a versatile order parameter in distinguishing
ground states \cite{PhysRevE.76.022101}. It
signals the phase boundaries shown in Fig. \ref{Phasediagramoftwosites}.
Remarkably, the phase diagram
found from the fidelity susceptibility for larger systems is the 
same as the one for $L=2$.

\begin{figure}[t!]
\begin{center}
\includegraphics[width=8.2cm]{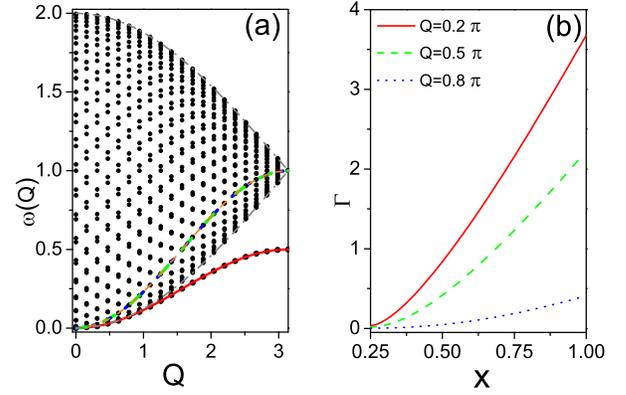}
\caption{(color online). (a) Energy spectra of 40-site spin-orbital
system at $x=y=1/4$. Dashed lines in the spin-orbital continuum
denote the spin, orbital and OBS excitation, all degenerate;
the (red) solid line below corresponds to the BS.
(b) The decay rate $\Gamma$ of the OBS for different momenta $Q$
with $y=x$ at $L\to\infty$.}
\label{lifetime1}
\end{center}
\end{figure}

{\it Excitations.---} In phase I of Fig.
\ref{Phasediagramoftwosites}, with boundaries given by:
$x+y=\frac12$, $x=-\frac14$ and $y=-\frac14$, the spins and orbitals are
fully polarized, and the ferro-ferro ground state $\vert 0 \rangle$ is
disentangled, i.e., can be factorized into spin and orbital sector.
It is now interesting to ask whether:
(i) the vanishing spin-orbital entanglement in the ground state implies
a suppression of joint spin-orbital quantum fluctuations,
and
(ii) collective spin-orbital excitations can form.
Using equation of motion method one finds spin (magnon) excitations
with dispersion $\omega_s(q)= (\frac{1}{4}+y)(1-\cos q)$, and
orbital (orbiton) excitations,
$\omega_t(q)= (\frac{1}{4}+x)(1-\cos q)$ \cite{Herzog}.
The stability of the orbitons (magnons) implies that $x>-\frac{1}{4}$
($y>-\frac{1}{4}$), and determines the QPT between phases I and II (IV),
respectively, while
the spin-orbital coupling only renormalizes the spectra.

For our purpose, it is straightforward to consider the propagation
of a pair of magnon and orbiton along the ferro-ferro chain, by
simultaneously exciting a single spin and a single orbital. The
translation symmetry imposes that total momentum $Q= 2 m \pi/L  $
$(m=0, \cdots, L-1)$ is conserved during scattering. The scattering
of magnon and orbiton with initial (final) momenta
$\{\frac{Q}{2}-q,\frac{Q}{2}+q\}$ ($\{\frac{Q}{2}-q',\frac{Q}{2}+q'\}$)
and total momentum $Q$
is represented by the Green's function \cite{Wortis},
\begin{eqnarray}
G(Q,\omega)= \frac{1}{L}\sum_{q,q'}\;\left\langle\left\langle
S_{\frac{Q}{2}-q'}^+T_{\frac{Q}{2}+q'}^+ | S_{\frac{Q}{2}-q}^-
T_{\frac{Q}{2}+q}^- \right\rangle\right\rangle,
 \label{Greenfunction}
\end{eqnarray}
for a combined spin ($S_{\frac{Q}{2}-q}^-$) and orbital
($T_{\frac{Q}{2}+q}^-$)
excitation. The spin-orbital continuum is given by
$\Omega(Q,q)=\omega_s(\frac{Q}{2}-q) + \omega_t (\frac{Q}{2}+q)$.
In the noninteracting case, the Green's
function exhibits square-root singularities at the edges of the
continuum \cite{Schneider}. Due to residual, attractive
interactions spin-orbital bound states (BSs) are shifted outside the
continuum \cite{Brink,Bala,Cojocaru}, see Fig. \ref{lifetime1}(a).
The collective mode is determined by
$1+  \frac{1}{2\pi} \int_{-\pi}^{\pi} dq  (\cos \frac{Q}{2}-
\cos q )^2/\left[\omega -\Omega(Q,q)\right]=0$.
The analytic solution of this equation is tedious but straightforward.
The collective BS with dispersion $\omega_{\textrm{BS}}(Q)$ is well
separated from the spinon-orbiton continuum [Fig. \ref{lifetime1}(a)]
at large $Q$. In the long-wave limit
the BS energy coincides with the Arovas-Auerbach line
\cite{Arovas}, i.e., the boundary of the continuum,
yet the BS remains undamped for $x+y>\frac12$.

\begin{figure}[t!]
\begin{center}
\includegraphics[width=7.4cm]{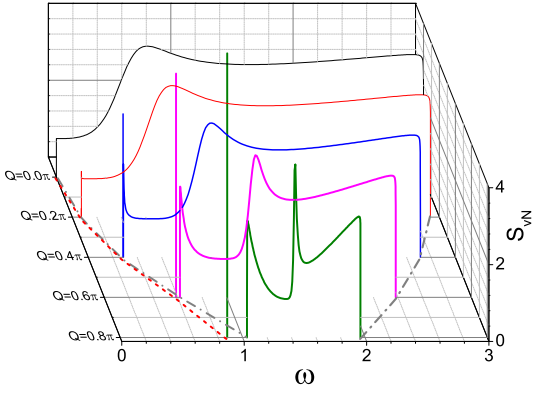}
 \caption{(color online). The vNE distribution of 400-site spin-orbital
system in subspace $P_\textrm{ST}=1$ for $x=y=1/2$ and different
momenta $Q$. Isolated vertical lines indicate the BS,
with dispersion given by the (red) dashed line.
The OBS in the center of spectra is damped.}
\label{SOE-N400}
\end{center}
\end{figure}

In addition, a collective mode of spin-orbital resonances,
\begin{eqnarray}
|\Psi(Q)\rangle=\frac{1}{\sqrt{L}} \sum_{m,l} a_l(Q) e^{iQm} S_{m}^-
T_{m+l}^-  \vert 0 \rangle,\label{wavefunction}
\end{eqnarray}
occurs inside the continuum.
Here $0 \le l \le L-1$ denotes the distance
between spin and orbital flips. Remarkably,
the spin and orbital flips are glued together at the same site with
$a_l(Q) =\delta_{l,0}$ at the SU(4) point \cite{Herzog}.
This coupled on-site BS (OBS) is a coherent superposition of local
modes, all of them with equal weight. It has dispersion
$\omega_{\textrm{OBS}}(Q)=  x+y - \frac{1}{2}\cos Q $,
which is degenerate with both $\omega_s(Q)$ and $\omega_t(Q)$
at $x=y=\frac14$, see Fig. \ref{lifetime1}(a).
This is reminiscent of the degeneracy of the three Goldstone modes
at the SU(4) point for $J=-1$ \cite{YouQuanLi,Sutherland}.
Moving away from the SU(4) point, the OBS decays due to residual
interactions into magnon-orbiton pairs, and the mean separation $\xi$
of spin and orbital excitations increases, i.e., $a_l(Q)\sim\exp(-l/\xi)$,
leading in the thermodynamic limit to a finite linewidth defined by
$\Gamma=\textrm{Im} G^{-1}(Q,\omega)$ \cite{decayrate}.
The decay rate of the
spin-orbital OBS increases with growing $x>\frac14$ and also for
decreasing momenta $Q$, as seen in Fig. \ref{lifetime1}(b).

{\it Entropy spectral function.--- }
To investigate the degree of entanglement of excited states,
we introduce
the vNE spectral function in the Lehmann representation,
\begin{eqnarray}
S_{\rm vN}(Q,\omega)=-\sum_n \textrm{Tr}\{\rho_s^{(\mu)} \log_2
\rho_s^{(\mu)}\}\delta\left\{\omega-\omega_n(Q)\right\},
\end{eqnarray}
where $(\mu)=(Q,\omega_n)$ denote momentum and excitation energy,
and $\rho^{(\mu)}_s=\textrm{Tr}_o|\Psi_n(Q)\rangle\langle\Psi_n(Q)|$
is obtained by tracing the orbital degrees of freedom. Let us
first consider the symmetric case, i.e., $x=y$. The Hilbert space
can be divided into two subspaces characterized by the parity
$P_\textrm{ ST}$ of the interchange of
$S \leftrightarrow T$, which is odd or even.
Translation symmetry
allows us to express the reduced density matrix $\rho_s$ in a
block-diagonal form, where each block corresponds to an irreducible
representation labeled by total momentum $Q$ and parity of exchange
symmetry $P_\textrm{ ST}$. The vNE can be obtained by diagonalizing
separately these blocks. In particular, the nondegenerate eigenstates
with odd parity can be explicitly cast in the form
$\frac{1}{\sqrt{2}} (S_{Q/2-q}^- T_{Q/2+q}^- - S_{Q/2+q}^-
T_{Q/2-q}^-) \vert 0 \rangle$. Consequently, the singlet-like
pair results in ${\cal S}_{\rm vN}=1$. For other spin-orbital
eigenstates with $P_\textrm{ ST}=1$, ${\cal S}_L\ge 1$, except the pure
spin and orbital waves.
Interestingly, we find that the parity is still conserved in subspace
$Q=0$ for $x \neq y$.
The strongly entangled spin-orbital BSs are reflected
by peaks in the von Neumann spectra $S_{\rm vN}(\omega)$,
shown in Fig. \ref{SOE-N400}. As momentum $Q$ decreases, the OBS-peak
in the center of spectra gets broader, implying a shorter lifetime.

\begin{figure}[t!]
\begin{center}
\includegraphics[width=8.0truecm]{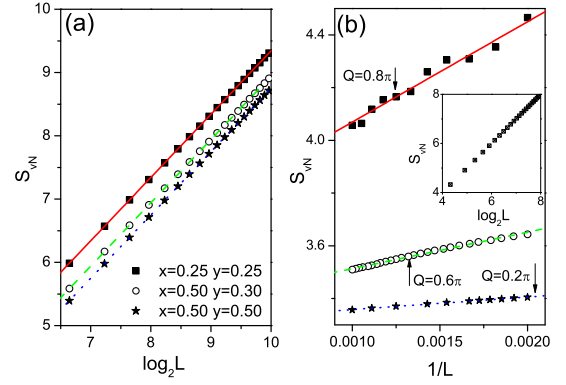}
 \caption{(color online). (a) Scaling behavior of entanglement entropy
${\cal S}_{\rm vN}$ of the spin-orbital BSs for $Q=0.8 \pi$. Lines
represent logarithmic fits ${\cal S}_\textrm{vN}= \log_2 L + c_0$, with
$c_0$ = $-0.659$, $-1.059$, $-1.251$, respectively.
(b) The scaling behavior of entanglement entropy of the OBS
for $x=y=1/2$. Lines are fitted by ${\cal S}_\textrm{vN}=c_1/L+c_0$,
with $c_0$ ($c_1$) = 3.69 (380.5), 3.37 (138.4) and 3.31 (47.6)
for $Q=0.8 \pi$, $0.6 \pi$ and $0.2 \pi$. The inset shows the
logarithmic behavior of ${\cal S}_{\rm vN}$ for the OBS with
$Q=0.8\pi$ and $x=y=1/4$.}
\label{Scaling1}
\end{center}
\end{figure}

Inspection of vNE spectra shows that the entanglement reaches a local
maximum at the BSs. Finite size scaling of vNE of spin-orbital BSs
reveals the asymptotic logarithmic scaling
${\cal S}_\textrm{vN}= \log_2 L + c_0$ shown in Fig. \ref{Scaling1}(a).
The same logarithmic scaling is found for the OBS at the SU(4) point
$x=y=\frac14$,
as seen in the inset of Fig. \ref{Scaling1}(b). However, far away from
the SU(4) point the scaling is entirely different and the entropy of the
OBS scales as a power law, ${\cal S}_\textrm{vN}=c_1/L+c_0$, as seen in
Fig. \ref{Scaling1}(b). This change of scaling from logarithmic to power
law in $1/L$ is controlled by the correlation length $\xi$ measuring the
average distance of spin and orbital excitations in the OBS wave
function (\ref{wavefunction}). From Eq. (\ref{wavefunction}) and
$a_l(Q)\sim\exp(-l/\xi)$ we obtain,
\begin{equation}
S_{\textrm{vN}}\simeq \log_2 \left\{L/(1+\xi)\right\},
\end{equation}
which yields $\log_2 L$ at $x=y=1/4$ where $\xi=0$. As $\xi$ increases
the correction to the vNE is $\propto  -\log_2 (1+\xi)$. Far away from
the SU(4) point, the OBS is damped and $\xi$ becomes extensive, i.e.,
$\xi/L\approx\tilde{c_0}-\tilde{c_1}/L$, and the vNE approaches a finite
value with a correction $\propto 1/L$ as shown in Fig. \ref{Scaling1}(b).
This close correspondence of the vNE of bound states and the correlation
length $\xi$ suggests to use the dynamic spin-orbital correlation
function as a probe of spin-orbital entanglement and as a qualitative
measure of the vNE spectra.

\begin{figure}[t!]
\begin{center}
\includegraphics[width=8.2cm]{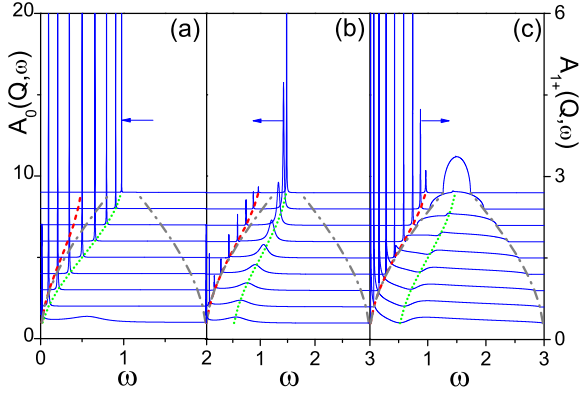}
\caption{(color online). The spectral function of the on-site excitation
$A_0(Q,\omega)$ for: (a) $x=y=1/4$, (b) $x=y=1/2$;
(c) the nearest-neighbor $A_{1+}(Q,\omega)$ for $x=y=1/2$.
The momenta range from $\pi/10$ (bottom) to $9 \pi/10$ (top); the peak
broadening is $\eta =0.01$. Dashed (red) and dotted (green) lines
correspond to the BS and OBS, while
gray dash-dot lines indicate the boundaries of the continuum.}
\label{A0kw}
\end{center}
\end{figure}

{\it Spectral functions.---}
Returning to TMOs, one realizes that joint spin-orbital excitations
are not created in the ferro-ferro ground state in
photoemission spectroscopy because of spin-conservation. On the
contrary, resonant inelastic x-ray scattering (RIXS)
\cite{Ishihara,Haverkort,Ament,Filomena,Sch12} is
in principle able to measure the spectral function of the coupled
spin-orbital excitations at distance $l$,
\begin{equation}
A_l(Q,\omega) =\frac{1}{\pi} \lim_{\eta \to 0} \textrm{Im} \left\langle
0 \left\vert \Gamma_Q^{(l)\dagger}\frac{1}{\omega+E_0 -H -i \eta}
\Gamma_Q^{(l)} \right\vert 0 \right\rangle.
\end{equation}
Here $\Gamma_Q^{(0)}=\frac{1}{\sqrt{L}}\sum_j e^{iQj} S_j^- T_j^-$
is the local excitation operator for an on-site spin-orbital excitation. 
We use as well
$\Gamma_Q^{(1_{\pm})}=
\frac{1}{\sqrt{2L}} \sum_j e^{iQj} (S_{j+1}^- \pm S_{j-1}^-) T_j^-$
for the nearest-neighbor excitation.
In the RIXS process an electron with spin up is excited by the incoming
x-rays from a deep-lying core level into the valence shell. For the time 
of its existence the core hole generates
a Coulomb potential and a strong spin-orbit coupling
that allows for the non-conservation of spin. Next the hole is
filled by an electron from the occupied valence band under the emission
of an x-ray. This RIXS process creates a joint spin-valence excitation
with momentum $Q_{\textrm{in}}-Q_{\textrm{out}}$ and energy
$\omega_{\textrm{in}}-\omega_{\textrm{out}}$,
which can unveil the spectral function of the spin-orbital excitation.

The on-site spectral function $A_{0}(Q,\omega)$ shown in Figs.
\ref{A0kw}(a,b) highlights the OBS. At the SU(4) point
[Fig. \ref{A0kw}(a)] it appears as a $\delta$-function,
$A_{0}(Q,\omega)=\delta\{\omega-\omega_{\textrm{OBS}}(Q)\}$,
whereas in Fig. \ref{A0kw}(b) the OBS is damped and its intensity
decreases strongly with $Q$. In the latter figure the
BS at the low energy side of the continuum
appears as weak additional feature, while
it is absent in (a), i.e., at the SU(4) point.
The nearest neighbor spectral function $A_{1+}(Q,\omega)$ in Fig.
\ref{A0kw}(c) shows both the spin-orbital continuum and the BS
outside of the continuum. Notably,
comparing with the vNE spectral function in Fig. \ref{SOE-N400}, we
find the same characteristic energies and similar intensity features
as in the RIXS spectra. The spectral function provides information
of various correlations, which are ingredients to derive the reduced
density matrices \cite{Peschel}.

{\it Summary.---}
In this Letter, we study a spin-orbital system and
extend the analysis of entanglement to excited states by
introducing the vNE spectral function. Our study demonstrates
that even in cases where the ground state of a spin-orbital chain
is fully disentangled, e.g., in the ferro-ferro state,
(i) the spin-orbital excitations are in general entangled,
(ii) maximal spin-orbital entanglement occurs for BSs which
appear as sharp peaks in the vNE spectra, and
(iii) the vNE of undamped BSs exhibits a logarithmic dependence on the
chain length $L$. We propose to study the dynamic spin-orbital
correlation function as a qualitative measure of the vNE spectra,
and suggest to use here RIXS as a promising technique.

W-L.Y. acknowledges support by the National Natural Science
Foundation of China (NSFC) under Grant No. 11004144. A.M.O.
acknowledges support by the Polish National Science Center
(NCN) under Project No. N202 069639.


\begin{thebibliography}{99}
\bibitem{decayrate} Dyson equation reads $G^{-1}(Q,\omega)=\omega
                    -\omega_{\textrm{OBS}}(Q)-\Sigma(Q,\omega)$,
                    with the total self-energy $\Sigma(Q,\omega)$.
                    The real part $\textrm{Re} G^{-1}(Q,\omega)$
                    identifies the effective energy pole, and the imaginary
                    part of self-energy is the inverse of lifetime $\tau$, 
                    i.e., $\Gamma=\textrm{Im}\Sigma(Q,\omega)$.
\end{thebibliography}
\end{document}